\documentclass{revtex4}
\usepackage{graphics,epsfig,color} 
\usepackage{amsmath,amssymb}
\usepackage[ansinew]{inputenc}

\begin{document}
\title{Scalar-tensor propagation of light in the inner solar system at the millimetric level}
\author{Olivier Minazzoli\footnote{also at ICRAnet, University of Nice Sophia Antipolis, 28 Avenue Valrose, 06103 Nice, France} and Bertrand Chauvineau}
\affiliation{UNS, OCA-ARTEMIS UMR 6162, Observatoire de la C\^ote d'Azur, Avenue Copernic, 06130 Grasse, France}
\begin{abstract}
In a recent paper \cite{MCPRD09}, motivated by forthcoming space experiments
involving propagation of light in the Solar System, we have proposed an extention of the IAU metric equations at the $c^{-4}$\ level in General Relativity. However, scalar-tensor theories may induce corrections numerically comparable to the $c^{-4}$ general relativistic terms.\ Accordingly, one first proposes in this paper an extension of \cite{MCPRD09}
to the scalar-tensor case. The case of a hierarchized system (such as the Solar system) is emphasized. In this case, the relevant metric solution is proposed. Then, the geodesic solution
relevant for propagation of light in the inner solar system at the
millimetric level is given in explicit form.

\end{abstract}
\maketitle

\section{Introduction}

Forthcoming space missions and missions in project -- such as LATOR \cite{lator}, TIPO \cite{tipo}, ASTROD \cite{astrod}, PLR \cite{PLR}, ODYSSEY \cite{odyssey} or SAGAS \cite{sagas} -- will require distance measurements at millimetric level in the Solar System.\ This corresponds to time transfer at the precision $%
10^{-11}\;s$. As argued in \cite{MCPRD09}, this requires a complete Solar
System metric at the $c^{-4}$\ level, in order to describe the laser links
involved in such experiments. This has been proposed in the framework of the
General Relativity (GR) theory in \cite{MCPRD09}, leading to an appropriate
extention of the metric equations recommanded by the IAU2000 resolution \cite{SKPetAJ03}.

The relative amplitude of the relativistic effects is of the order of $\epsilon
=GM/rc^{2}$, where $M$ and $r$ are some characteristic mass and distance.
The so-called first order terms are of order $\epsilon $, the second order
terms of order $\epsilon ^{2}$. In the inner Solar System, $\epsilon $ is
typically of the order $10^{-8}$, and can be sensitively greater ($10^{-7}$)
for photons entering well inside Mercury's orbit, and can even be as large
as $10^{-6}$ for photons grazing the Sun.

On the other hand, there is a surge of interest in scalar-tensor (ST) theories since about two
decades.\ Indeed, the gravitational sector of a lot of tentative fundamental
theories, like string or (modern) Kaluza-Klein, turns to be described by a
metric tensor plus a scalar field (Brans-Dicke \cite{FMbook03,Fbook04} or not \cite{CBPrD07}).
Besides, alternative theories to GR (including ST) are also sometimes required by some authors to deal
with the so-called dark energy (cosmological level), dark matter (galactic
level), Pioneer or fly-by anomalies (solar system level) problems. Phenomenologically, the  divergence between ST theories and GR is quantified by the Post-Newtonian (PN) $\gamma $
factor ($=1$ in GR) entering the $c^{-2}$ term in the space-space components
of the metric tensor. From the present observations, $\left|
\gamma -1\right| $ can at best reach values of the order of $10^{-5}$
\cite{WLivRev06}.\ But, some theoretical considerations strongly suggest
$\gamma$ could have been driven from any ''initial'' value to a value close to unity by the
cosmological expansion (more precisely, ST theories are driven to GR, as
soon as these theories fulfil some (not very constraining) conditions), and
even claim the $1-\gamma$ current value should be of order $10^{-7}$ or $10^{-8}$ \cite{DNPRL93,DNPRD93}. On the other hand, $c^{-4}$ terms are
typically $10^{6}$ to $10^{8}$\ smaller than $c^{-2}$ terms. All these
reasons make relevant the extension of the gravitational framework proposed
in \cite{MCPRD09} to encompass the ST case, as much as $c^{-4}$ space-space metric terms have to be taken into account in light propagation
problems.

In principle, this would require new definitions of multipolar moments at the $c^{-2}$ level. But other publications proposed  $c^{-2}$ multipolar moments in ST theories \cite{KVPR04} or in a parametrized post-newtonian framework \cite{KSPrD00}. Thus we do not discuss this point in this paper and focus only on the $c^{-4}$ metric side of the problem. Another reason to discard this point is that non-monopolar terms turn out to have numerically negligible contributions for links we are interested in in this paper (inner Solar system case).

In section \ref{sec:term}, one defines a terminology relevant to the considered problem.
Since the Einstein conformal representation plays a central role in the
approach followed in this paper, the section \ref{sec:EivsJo} is devoted to the conformal
link between the representations of ST theories, and to the related
notations we will use. Section \ref{sec:TFE} is dedicated to the derivation of the ST
field equations up to $O\left( c^{-5}\right) $ terms. In section \ref{sec:omega0}, these
field equations are rewritten up to $O\left( c^{-5},\omega
_{0}^{-1}c^{-4}\right) $ terms, for applications taking explicitely our
present experimental knowledge on gravity into account. Considering
applications to Solar System-like systems, one defines hierarchized
systems in section \ref{sec:hierarchy}.\ This is made by defining a quantity $\mu $ that quantifies how much
the system is gravitationally dominated by its most massive body.
Accordingly, the field equations are rewritten up to $O\left( c^{-5},\omega
_{0}^{-1}c^{-4},\mu c^{-4}\right) $ terms. In this case, the explicit metric solution that can
be used in relevant applications is written in well-suited coordinates. Finally, in section VII, we give the explicit form of isotropic
geodesics relevant for time transfer and ranging problems in the inner solar system at
the millimetric level of accuracy (required by forthcoming space missions or
missions in project).

\section{Terminology : definition of the PN/BM and PN/RM metrics}
\label{sec:term}

The PN approximation is based on the assumption of a weak gravitational
field and low velocities for both the sources and the
(test) body (ie. velocities of the order $\sqrt{GM/r}$ or less, $M$
being some caracteristic mass of the system). It formally consists in looking for solutions under the form of
an expansion in powers of $1/c$. The usually so-called $n$PN order terms in the metric, leading to $%
c^{-2n}$ terms in the equation of motion of a body describing a bounded orbit, are terms
of orders $c^{-2n-2}$ in $g_{00}$, $c^{-2n-1}$ in $g_{0i}$ and $c^{-2n}$ in $%
g_{ij}$. In this case, the Ricci tensor components have to be developped the same way as (18) in \cite{MCPRD09}. In the present paper, a metric developped this way will be refered as the $n$PN/BM metric (BM meaning "Bounded Motion" \ for test particles). It is particularly well-adapted for
studying bounded motions in systems made by non-relativistic massive bodies, as the
Solar System is. 

However, since we are interested in the propagation of light, we are led to relax the hypothesis on the velocity of the test particle whose motion is considered. Of course, this doesn't change the metric, but the terms to be
considered in the metric components are not the same as in the PN/BM
problem.\ Indeed, the terms leading to $c^{-2n}$ terms in the equation of
motion of a test particle moving at relativistic velocity  are terms of
order $c^{-2n}$ in both $g_{00}$, $g_{0i}$
and $g_{ij}$. In this case, the Ricci tensor components have to be developped the same way as (19) in \cite{MCPRD09}. In the present paper, a metric developped this way will be refered as the $n$PN/RM
metric (RM meaning "Relativistic Motion" \ for test particles). A PN/RM
metric is particularly well-suited for studying \textit{relativistic} motions of test bodies (for instance, the propagation of light) in systems made by non-relativistic bodies, as the Solar System is. 


The present paper deals with the PN/RM problem since we are concerned in propagation of light.

\section{The ST theories in Einstein vs Jordan representations}
\label{sec:EivsJo}

The Jordan representation of the ST theories is described by the action 
\begin{eqnarray}
S&=&\frac{c^{4}}{16\pi G}\int d^{4}x\sqrt{-g}\left[ \Phi R-\frac{\omega \left(
\Phi \right) }{\Phi }g^{\alpha \beta }\partial _{\alpha }\Phi \partial
_{\beta }\Phi \right] +\int d^{4}x\sqrt{-g}L_{NG}\left( \Psi ,g_{\mu \nu
}\right) .  \label{jrepr}
\end{eqnarray}
In this representation, the gravitational sector of the theory is described
by the Jordan metric $g_{\alpha \beta }$ and the scalar field $\Phi $, while
the non-gravitational fields are symbolically represented by $\Psi $. The
scalar field couples directly with the metric, leading to rather complicated
field equations. Besides, the kinetic term associated to the scalar field
doesn't have the standard form, involving a scalar field function $\omega $,
the function characterizing the ST theory we are dealing with. On the other
hand, the non-gravitational lagrangian $L_{NG}$ doesn't depend on the scalar
field, leading to simple equations of motion ($\nabla _{\alpha }T^{\alpha
\beta }=0$), with the nice consequence that the weak equivalence principle
applies in this representation of the theory.

To overcome the just mentionned drawbacks, one could be tempted to resort to
a dependent variables change 
\begin{equation*}
\left( g_{\alpha \beta },\Phi \right) \longrightarrow \left( \overset{\_}{g}%
_{\alpha \beta },\varphi \right)
\end{equation*}
chosen in such a way that the action (\ref{jrepr}) transforms into 
\begin{eqnarray}
S&=&\frac{c^{4}}{16\pi G}\int d^{4}x\sqrt{-\overset{\_}{g}}\left[ \overset{\_}{%
R}-2\overset{\_}{g}^{\alpha \beta }\partial _{\alpha }\varphi \partial
_{\beta }\varphi \right] +\int d^{4}x\sqrt{-\overset{\_}{g}}\overset{\_}{L}_{NG}  \label{erepr}
\end{eqnarray}
$\overset{\_}{L}_{NG}$ depending on $\Psi $, $\overset{\_}{g}_{\mu \nu }$
and $\varphi $ in a way to be precised later ($\overset{\_}{R}$ and $%
\overset{\_}{g}$\ correspond to $R$\ and $g$, but with $g_{\alpha \beta }$\
replaced by $\overset{\_}{g}_{\alpha \beta }$). Since the scalar field
doesn't couple with the metric, (\ref{erepr}) is refered to as the Einstein
representation of the theory. 

The form (\ref{erepr}) is achieved by
considering a conformal transformation of the metric 
\begin{equation}
g_{\alpha \beta }=A\left( \varphi \right) ^{2}\overset{\_}{g}_{\alpha \beta
}.  \label{transfoconf}
\end{equation}
From the induced transformation of the Ricci scalar \cite{Wbook84}, and up to
a divergence term, (\ref{erepr}) turns into 
\begin{eqnarray}
S&=&\frac{c^{4}}{16\pi G}\int d^{4}x\sqrt{-g}\left[ \frac{1}{A^{2}}R+\left\{ \frac{6}{A^{4}}-\frac{2}{A^{2}}\left( \frac{d\varphi }{dA}\right)
^{2}\right\} g^{\alpha \beta }\partial _{\alpha }A\partial _{\beta }A\right] +\int d^{4}x\sqrt{-g}A^{-4}\overset{\_}{L}_{NG}.  \label{jfrome}
\end{eqnarray}
Comparing (\ref{jfrome}) with (\ref{jrepr}) suggests :

- the link between the Jordan and Einstein representations of the scalar
field 
\begin{equation}
\Phi =\frac{1}{A\left( \varphi \right) ^{2}};  \label{Adef}
\end{equation}

- the identification 
\begin{equation}
\frac{\omega \left( \Phi \right) }{\Phi }g^{\alpha \beta }\partial _{\alpha
}\Phi \partial _{\beta }\Phi =\left\{ \frac{2}{A^{2}}\left( \frac{d\varphi }{%
dA}\right) ^{2}-\frac{6}{A^{4}}\right\} g^{\alpha \beta }\partial _{\alpha
}A\partial _{\beta }A;  \label{stfctslink}
\end{equation}

- the Einstein representation of the non gravitational lagrangian 
\begin{eqnarray*}
\overset{\_}{L}_{NG}&=&A\left( \varphi \right) ^{4}L_{NG}\left( \Psi ,g_{\mu
\nu }\right) \\
&=&A\left( \varphi \right) ^{4}L_{NG}\left( \Psi ,A\left( \varphi
\right) ^{2}\overset{\_}{g}_{\mu \nu }\right) .
\end{eqnarray*}
(\ref{Adef}) and (\ref{stfctslink}) lead to the link between the functions $\omega \left(
\Phi \right) $ and $A\left( \varphi \right) $ (equivalently characterizing
the considered ST theory) 
\begin{equation}
\left( 2\omega +3\right) \left( \frac{d\ln \left| A\right| }{d\varphi }%
\right) ^{2}=1.  \label{sclink}
\end{equation}
(\ref{sclink}) requires $\omega >-3/2$.\ This results from the fact one has
imposed the sign of the scalar kinetic energy term in Einstein
representation (\ref{erepr}) in order to ensure the dynamical stability of
the theory \cite{WLNP07}.

The link between the stress tensor components in the two representations
follows from the general stress tensor definition 
\begin{eqnarray*}
\delta g^{\alpha \beta } &\longrightarrow &\delta
\int d^{4}x\sqrt{-g}L_{NG}\equiv -\frac{1}{2}\int d^{4}x\sqrt{-g}T_{\alpha
\beta }\delta g^{\alpha \beta } \\
\delta \overset{\_}{g}^{\alpha \beta } &\longrightarrow &\delta \int d^{4}x\sqrt{-\overset{\_}{g}}\overset{\_}{L}_{NG}\equiv -%
\frac{1}{2}\int d^{4}x\sqrt{-\overset{\_}{g}}\overset{\_}{T}_{\alpha \beta
}\delta \overset{\_}{g}^{\alpha \beta }.
\end{eqnarray*}
Since $\sqrt{-\overset{\_}{g}}\overset{\_}{L}_{NG}=\sqrt{-g}L_{NG}$, and
since the scalar field is not varied in this metric variation process (no ambiguity since the
two versions $\varphi $ and $\Phi $\ of the scalar field are related in the
non metric dependent way (\ref{Adef})), it directly turns that 
\begin{equation*}
\overset{\_}{T}_{\alpha \beta }=A^{2}T_{\alpha \beta }.
\end{equation*}
For the mixed and contravariant components, it follows 
\begin{equation*}
\overset{\_}{T}_{\alpha }^{\beta }=A^{4}T_{\alpha }^{\beta }\text{ \ \ ( }%
\Longrightarrow \;\overset{\_}{T}=A^{4}T\text{ ), \ \ }\overset{\_}{T}%
^{\alpha \beta }=A^{6}T^{\alpha \beta }
\end{equation*}
the indexes being raised/lowered by the metric involved in the corresponding
representation.

\bigskip

\textit{Eliminating }$A$

\bigskip

It is clear $A$ can be eliminated between the two representations of the
scalar field using (\ref{Adef}).\ This way, the considered ST theory is
represented by the function $\Phi \left( \varphi \right) $ in Einstein
representation. From (\ref{sclink}), this function is linked to $\omega
\left( \Phi \right) $ by 
\begin{equation}
\left( 2\omega +3\right) \left( \frac{d\ln \Phi }{d\varphi }\right)
^{2}=4.  \label{sclink2}
\end{equation}
The conformal transformation (\ref{transfoconf}), the link between the non
gravitational lagrangians and stress tensors now write 
\begin{equation}
\overset{\_}{g}_{\alpha \beta }=\Phi  g_{\alpha \beta }
\label{transfoconf2}
\end{equation}
\begin{equation*}
\overset{\_}{L}_{NG}=\Phi ^{-2}L_{NG}\left( \Psi ,\Phi ^{-1}\overset{\_}{g}%
_{\mu \nu }\right)
\end{equation*}
\begin{eqnarray}
\overset{\_}{T}_{\alpha \beta }&=&\Phi ^{-1}T_{\alpha \beta },\notag\\
\overset{\_}{T}_{\alpha }^{\beta }&=&\Phi ^{-2}T_{\alpha }^{\beta }(\Longrightarrow \;\overset{\_}{T}=\Phi ^{-2}T) \notag\\
\overset{\_}{T}^{\alpha \beta }&=&\Phi ^{-3}T^{\alpha \beta }.  \label{transfostress}
\end{eqnarray}

As it turns from equation (\ref{sclink2}), $\varphi $ is defined up to
a sign and an additive constant.\ When needed in the following, the sign
will be fixed by the choice 
\begin{equation}
\sqrt{2\omega +3}\frac{d\ln \Phi }{d\varphi }=2.  \label{sclink3}
\end{equation}
\section{The field equations}
\label{sec:TFE}

\subsection{Using the Einstein representation gravitational
field variables}

From (\ref{erepr}), the field equations can be written 
\begin{eqnarray}
\overset{\_}{R}^{\alpha \beta } &=&\frac{8\pi G}{c^{4}}\left( \overset{\_}{T}%
^{\alpha \beta }-\frac{1}{2}\overset{\_}{T}\overset{\_}{g}^{\alpha \beta
}\right) +2\overset{\_}{\partial }^{\alpha }\varphi \overset{\_}{\partial }%
^{\beta }\varphi   \notag \\
\partial _{\alpha }\left( \sqrt{-\overset{\_}{g}}\overset{\_}{\partial }%
^{\alpha }\varphi \right) &=&\frac{2\pi G}{c^{4}}\overset{\_}{T}\sqrt{-%
\overset{\_}{g}}\frac{d\ln \Phi }{d\varphi } \label{fieldeq}
\end{eqnarray}
the function $\Phi \left( \varphi \right) $\ characterizing the ST theory
explicitly entering the scalar field equation. As usual in PN
approximation, let us write the scalar field as 
\begin{equation}
\varphi =\varphi _{0}+\frac{\overset{\left( 2\right) }{\varphi }}{c^{2}}+%
\frac{\overset{\left( 4\right) }{\varphi }}{c^{4}}+O\left( c^{-5}\right)
\label{escdevel}
\end{equation}
where $\varphi _{0}$\ is constant and $\overset{\left( 2\right) }{\varphi }$ and $\overset{\left( 4\right) }{\varphi }$ are zeroth order terms. (Remark that, since $\varphi $ is defined
up to an additive constant, it is not restrictive to set $\varphi _{0}=0$.) 
As a consequence, it turns out that, under the standard PN assumptions, $\overset{\_}{R}^{ij}=O\left( c^{-4}\right) $, so that the Strong Spatial Isotropy Condition (SSIC) \cite{DSXPRD91} applies in this representation.\ It is then possible to choose a coordinate system in which
the Einstein metric takes the following form at the 2PN/RM level \cite{MCPRD09}
\begin{eqnarray}
\overset{\_}{g}_{00} &=&-1+\frac{2w}{c^{2}}-\frac{2w^{2}}{c^{4}}+O\left(
c^{-5}\right)  \label{emetric} \\
\overset{\_}{g}_{0i} &=&-\frac{4w_{i}}{c^{3}}+O\left( c^{-5}\right)  \notag
\\
\overset{\_}{g}_{ij} &=&\delta _{ij}\left( 1+\frac{2w}{c^{2}}+\frac{2w^{2}}{%
c^{4}}\right) +\frac{4\tau _{ij}}{c^{4}}+O\left( c^{-5}\right) .  \notag
\end{eqnarray}
Putting $\Phi(\varphi_{0})=1$ is not restrictive since $\Phi $ enters (\ref
{fieldeq}) throught its logarithm derivative. Hence, the scalar field function develops as 
\begin{equation}
\Phi \left( \varphi \right) =1+\frac{1}{c^{2}}\Phi _{0}^{\prime }\overset{%
\left( 2\right) }{\varphi }+\frac{1}{c^{4}}\left( \Phi _{0}^{\prime }%
\overset{\left( 4\right) }{\varphi }+\frac{1}{2}\Phi _{0}^{\prime \prime }%
\overset{\left( 2\right) }{\varphi }^{2}\right) +O\left( c^{-5}\right)
\label{jscdevel}
\end{equation}
where $\Phi _{0}^{\prime }$ and $\Phi _{0}^{\prime
\prime }$\ stand for the values of the derivatives of $\Phi $ at $\varphi
_{0}$.\ Setting 
\begin{eqnarray}
\sigma &=&\frac{1}{c^{2}}\left( T^{00}+T^{kk}\right)  \label{sources1} \\
\sigma ^{i} &=&\frac{1}{c}T^{0i}  \notag \\
\sigma ^{ij} &=&T^{ij}-T^{kk}\delta _{ij}\text{ \ \ \ \ \ ( }\Longrightarrow
\;\sigma ^{kk}=-2T^{kk}\text{\ )}  \notag
\end{eqnarray}
(which, from standard PN assumptions, are $c^{0}$ order quantities) and
using (\ref{transfostress}), the $\left( 00\right) $, $\left( 0i\right) $
and $\left( ij\right) $\ field equations (\ref{fieldeq}) lead respectively
to 
\begin{equation}
\triangle w+\frac{1}{c^{2}}\left( 3\partial _{tt}w+4\partial
_{tk}w_{k}\right) +\frac{3}{c^{2}}\Phi _{0}^{\prime }\overset{\left(
2\right) }{\varphi }\triangle w=-4\pi G\sigma +O\left( c^{-3}\right)
\label{eg00}
\end{equation}
\begin{equation}
\triangle w_{i}-\partial _{ik}w_{k}-\partial _{ti}w=-4\pi G\sigma
^{i}+O\left( c^{-2}\right)  \label{eg0i}
\end{equation}
\begin{eqnarray}
\Theta _{ij}\left( \tau _{kl}\right) &=&\partial _{i}w\partial
_{j}w-\partial _{t}\left( \partial _{i}w_{j}+\partial _{j}w_{i}\right)-2\delta _{ij}\partial _{t}\left( \partial _{t}w+\partial _{k}w_{k}\right)+4\pi G\sigma ^{ij}+\partial _{i}\overset{\left( 2\right) }{\varphi }%
\partial _{j}\overset{\left( 2\right) }{\varphi }+O\left( c^{-1}\right) 
 \label{egij}
\end{eqnarray}
where $\Theta _{ij}$ is defined, as in \cite{MCPRD09}, by
\begin{eqnarray}
\Theta _{ij}\left( \tau _{kl}\right) &\equiv &\partial
_{ik}\tau _{jk}+\partial _{jk}\tau _{ik}-\triangle \tau _{ij}-\partial
_{ij}\tau _{kk}.  \notag
\end{eqnarray}
The scalar field equation gives 
\begin{eqnarray}
\triangle \overset{\left( 2\right) }{\varphi }+\frac{1}{c^{2}}\left(
-\partial _{tt}\overset{\left( 2\right) }{\varphi }+\triangle \overset{%
\left( 4\right) }{\varphi }\right) 
+\frac{1}{c^{2}}\left( 4\Phi _{0}^{\prime
}-\frac{\Phi _{0}^{\prime \prime }}{\Phi _{0}^{\prime }}\right) \overset{%
\left( 2\right) }{\varphi }\triangle \overset{\left( 2\right) }{\varphi } 
=-2\pi G\Phi _{0}^{\prime }\left( \sigma +\frac{\sigma ^{kk}}{c^{2}}\right)
+O\left( c^{-3}\right) .  \label{esc}
\end{eqnarray}
Remark that, in contrast to the GR case, the (00) equation is not linear,
because of the $c^{-2}\overset{\left( 2\right) }{\varphi }\triangle w$\
term. The scalar field equation also contains a non-linear $c^{-2}$\ term.

Now, combining (\ref{eg00}) and (\ref{esc}) leads to 
\begin{equation*}
\triangle \left( \overset{\left( 2\right) }{\varphi }-\frac{1}{2}\Phi
_{0}^{\prime }w\right) =O\left( c^{-2}\right) .
\end{equation*}
Accordingly, let us choose
\begin{equation}
\overset{\left( 2\right) }{\varphi }=\frac{1}{2}\Phi _{0}^{\prime }w.
\label{solphi2}
\end{equation}
Hence, defining $\chi \equiv \overset{\left( 4\right) }{\varphi }/\Phi
_{0}^{\prime }$ 
\begin{equation}
\varphi =\varphi _{0}+\frac{\Phi _{0}^{\prime }}{2c^{2}}w+\frac{\Phi
_{0}^{\prime }}{c^{4}}\chi +O\left( c^{-5}\right) .  \label{escdevel2}
\end{equation}
The metric field variables $w$, $w_{i}$ and $\tau _{ij}$ are now decoupled
from the scalar field $\chi $. The system constraining $w$, $w_{i}$
and $\tau _{ij}$ now writes 
\begin{equation}
\triangle w+\frac{1}{c^{2}}\left( 3\partial _{tt}w+4\partial
_{tk}w_{k}\right) +\frac{3}{2c^{2}}\Phi _{0}^{\prime 2}w\triangle w=-4\pi
G\sigma +O\left( c^{-3}\right)  \label{eg00a}
\end{equation}
\begin{equation}
\triangle w_{i}-\partial _{ik}w_{k}-\partial _{ti}w=-4\pi G\sigma
^{i}+O\left( c^{-2}\right)  \label{eg0ia}
\end{equation}
\begin{eqnarray}
\Theta _{ij}\left( \tau _{kl}\right) =-\partial _{ti}w_{j}-\partial
_{tj}w_{i}\label{egija}
+\left( 1+\frac{1}{4}\Phi _{0}^{\prime 2}\right) \partial
_{i}w\partial _{j}w
-2\delta _{ij}\left( \partial _{tt}w+\partial
_{tk}w_{k}\right) +4\pi G\sigma ^{ij}+O\left( c^{-1}\right)  
\end{eqnarray}
$\chi $ being obtained in a second step, by solving 
\begin{eqnarray}
\triangle \chi -2\left( \partial _{tt}w+\partial _{tk}w_{k}\right)\label{esca} +\frac{1}{%
4}\left( \Phi _{0}^{\prime 2}-\Phi _{0}^{\prime \prime }\right) w\triangle
w=-2\pi G\sigma ^{kk}+O\left( c^{-1}\right) .   
\end{eqnarray}
Remark that $\triangle w$\ may be replaced by $-4\pi G\sigma $\ in the non
linear terms of equations (\ref{eg00a}) and (\ref{esca}).

\subsection{Back to Jordan representation}

Let us use the function $\omega \left( \Phi \right) $ and its derivative $%
\omega ^{\prime }\left( \Phi \right) $\ instead of $\Phi ^{\prime }\left(
\varphi \right) $ and $\Phi ^{\prime \prime }\left( \varphi \right) $. One
finds, using (\ref{sclink3})

\begin{eqnarray*}
\Phi _{0}^{\prime } &=&\frac{2}{\sqrt{2\omega _{0}+3}} \\
\Phi _{0}^{\prime \prime } &=&\frac{4}{2\omega _{0}+3}\left( 1-\frac{\omega
_{0}^{\prime }}{2\omega _{0}+3}\right) =\Phi _{0}^{\prime 2}-\frac{4\omega
_{0}^{\prime }}{\left( 2\omega _{0}+3\right) ^{2}}.
\end{eqnarray*}
One now goes back to Jordan representation using (\ref{transfoconf2}), with,
from (\ref{jscdevel}) and (\ref{escdevel2}),
\begin{eqnarray}
\Phi ^{-1}=1-\frac{2w}{c^{2}\left( 2\omega _{0}+3\right) }\label{eq:phim1}
+\frac{1}{%
c^{4}\left( 2\omega _{0}+3\right) }\left[ \frac{2}{2\omega _{0}+3}\left( 1+%
\frac{\omega _{0}^{\prime }}{2\omega _{0}+3}\right) w^{2}-4\chi \right] 
+O\left( c^{-5}\right) . 
\end{eqnarray}
Now let us put 
\begin{eqnarray*}
\gamma &=&\frac{\omega _{0}+1}{\omega _{0}+2}\\
\beta &=&1+\frac{\omega _{0}^{\prime }}{\left( 2\omega _{0}+3\right) \left( 2\omega
_{0}+4\right) ^{2}}\\
G_{eff}&=&\frac{2\omega _{0}+4}{2\omega_{0}+3}G
\end{eqnarray*}
and let us define 
\begin{equation*}
\left( U,U_{i},U_{ij},P\right) =\frac{2\omega _{0}+4}{2\omega _{0}+3}\left(
w,w_{i},\tau _{ij},\chi \right)
\end{equation*}
and the related quantities $\left( W,W_{i},W_{ij}\right) $ by 
\begin{eqnarray*}
W&=&U+\left( 1-\gamma \right) \frac{P}{c^{2}}\\
W_{i}&=&U_{i}\\
W_{ij}&=&U_{ij}-\left( 1-\gamma \right) P\delta _{ij}.
\end{eqnarray*}
Using (\ref{emetric}) and (\ref{eq:phim1}), one gets the Jordan metric 
\begin{eqnarray}
g_{00} &=&-1+\frac{2W}{c^{2}}-\beta \frac{2W^{2}}{c^{4}}+O\left(
c^{-5}\right)  \label{jmetric} \\
g_{0i} &=&-\left( \gamma +1\right) \frac{2W_{i}}{c^{3}}+O\left( c^{-5}\right)
\notag \\
g_{ij} &=&\delta _{ij}\left\{ 1+\gamma \frac{2W}{c^{2}}+\left( \gamma
^{2}+\beta -1\right) \frac{2W^{2}}{c^{4}}\right\}
 +\left( \gamma +1\right) 
\frac{2W_{ij}}{c^{4}}+O\left( c^{-5}\right) .  \notag
\end{eqnarray}
where the functions $\left( W,W_{i},W_{ij},P\right) $\ satisfy the following field equations -- after some algebra from (\ref
{eg00a}-\ref{esca}) and (\ref{sources1}) 
\begin{eqnarray}
&&\square W+\frac{1+2\beta -3\gamma }{c^{2}}W\triangle W+\frac{2}{c^{2}}\left(
1+\gamma \right) \partial _{t}J =-4\pi G_{eff}\Sigma +O\left( c^{-3}\right)\notag \\
&&\triangle W_{i}-\partial _{i}J =-4\pi G_{eff}\Sigma ^{i}+O\left(
c^{-2}\right)  \notag \\
&&\triangle W_{ij}+\partial _{i}W\partial _{j}W+2\left( 1-\beta \right) \delta
_{ij}W\triangle W-\partial _{i}J_{j}-\partial _{j}J_{i}-2\gamma \delta
_{ij}\partial _{t}J=-4\pi G_{eff}\Sigma ^{ij}+O\left( c^{-1}\right) 
\notag \\
&&\triangle P+2\frac{\beta -1}{1-\gamma }W\triangle W-2\partial _{t}J  =-4\pi G_{eff}\frac{\Sigma ^{kk}}{3\gamma -1}+O\left( c^{-1}\right) . \label{fieldeq2} 
\end{eqnarray}
In (\ref{fieldeq2}), one has set 
\begin{eqnarray}
J &=&\partial _{t}U+\partial _{k}U_{k} \label{eq:jaugeJ} \\
&=&\partial _{t}W+\partial _{k}W_{k}+O\left( c^{-2}\right) \notag \\
J_{i} &=&\partial _{k}U_{ik}-\frac{1}{2}\partial _{i}U_{kk}+\partial
_{t}U_{i} \notag\\
&=&\partial _{k}W_{ik}-\frac{1}{2}\partial _{i}W_{kk}+\partial _{t}W_{i}-%
\frac{1-\gamma }{2}\partial _{i}P \label{eq:ji} 
\end{eqnarray}
and, for the matter part of the equations
\begin{eqnarray*}
&&\Sigma =\frac{1}{c^{2}}\left( T^{00}+\gamma T^{kk}\right) \\
&&\Sigma ^{i} =\frac{1}{c}T^{0i} \\
&&\Sigma ^{ij} =T^{ij}-\gamma T^{kk}\delta _{ij}\text{ \ \ \  (}%
\Longrightarrow \;\Sigma ^{kk}=-\left( 3\gamma -1\right) T^{kk}\text{)}
\end{eqnarray*}
Let us remark that the quantity $2\left( \beta -1\right) /\left( 1-\gamma
\right) $ ($=\omega _{0}^{\prime }\left( 2\omega _{0}+3\right) ^{-1}\left(
2\omega _{0}+4\right) ^{-1}$) is not diverging, even if $\gamma $\ is
(arbitrarily) close to unity. Besides, no new PN parameter appears neither in the $c^{-4}$ space-space part of the metric nor in the corresponding field equations, as stressed in \cite{DFPRD96}. 

This form is relevant in all sufficiently weak gravitational field, even in
systems where the ST theory is not very close to GR, i.e. where the PN parameters $%
\gamma $ and $\beta $ are not close to unity. A priori, this may occur even if $\gamma $ and $\beta $ are close to unity in some (other)
regions of the universe, as in the Solar System, as soon as the ST theory is
not (in some sense) close to the Brans-Dicke one (in Brans-Dicke gravity, $\omega $ doesn't depend
on the scalar field, so that it has the same value in all the space-time regions of the universe).

Let us point out that the numerical values
of the coefficients $\gamma ^{2}+\beta -1$ and $\gamma +1$\ entering $g_{ij}$
in (29) can be chosen independently one to another, since both $\beta $ and $%
\gamma $ enter these coefficients.\ Hence, there is no a priori relation
between the coefficients of the $c^{-4}$-terms $W^{2}$ and $W_{ij}$,
contrary to what may be suggested by the form of the metric chosen in \cite{KZ} (in the one-mass case). More precisely, this doesn't mean
the form chosen by \cite{KZ} is uncorrect, but rather that
this form doesn't encompass the (general) ST case
(but it encompasses the GR case, as it must be).

\subsection{Harmonic gauges}

Since the use of the harmonic gauge (HG) is recommended by the IAU, let us
consider the field equations in this gauge.\ Of course one has to
specify the representation in which the HG is prescribed. The Jordan HG
condition reads 
\begin{equation*}
g^{\alpha \beta }\Gamma _{\alpha \beta }^{\sigma }=0
\end{equation*}
and it leads to, for the space ($\sigma =k$) component 
\begin{equation}
\left( \gamma -1\right) \partial _{k}U=O\left( c^{-2}\right) .
\label{STJHGk}
\end{equation}
As expected from known results in GR \cite{DSXPRD91}, this condition reduces to a triviality in the case $\gamma
=1 $.\ On the other hand, if $\gamma \neq 1$, (\ref{STJHGk}) shows that the coordinate system in which the metric takes the (Jordan) form
(\ref{jmetric}), corresponding to SSIC in Einstein representation, doesn't
encompass (Jordan) harmonic coordinates in the ST case. In other terms,
(Jordan) harmonic coordinates are incompatible with the SSIC in Einstein
representation.

One could rather choose to impose the HG condition on the metric in Einstein
representation 
\begin{equation*}
\overset{\_}{g}^{\alpha \beta }\overset{\_}{\Gamma }_{\alpha \beta }^{\sigma
}=0
\end{equation*}
since the Einstein metric (\ref{emetric})\ satisfies the SSIC. From (\ref{emetric}), this means
one imposes $w$, $w_{i}$ and $\tau _{ij}$ to satisfy
\begin{eqnarray*}
\partial _{t}w+\partial _{k}w_{k} &=&O\left( c^{-2}\right) \\
\partial _{k}\tau _{ik}-\frac{1}{2}\partial _{i}\tau _{kk}+\partial
_{t}w_{i} &=&O\left( c^{-1}\right) .
\end{eqnarray*}
Translated in terms of $\left( U,U_{i},U_{ij}\right) $, this takes exactly
the same form, i.e., using (\ref{eq:jaugeJ}-\ref{eq:ji})
\begin{equation}
J=O\left( c^{-2}\right) \text{ \ \ , \ \ }J_{i}=O\left( c^{-1}\right) .
\label{eharmo}
\end{equation}
It turns out this corresponds to the Nutku gauge constraints \cite{KVPR04,XNDetASR09}, meaning that imposing the HG in the Einstein representation is equivalent to impose the Nutku gauge in the Jordan representation.
Using (\ref{eharmo}), the three first equations of (\ref{fieldeq2}) take the reduced form 
\begin{eqnarray}
&&\square W+\frac{1+2\beta -3\gamma }{c^{2}}W\triangle W \notag \\
&&~~~~=-4\pi G_{eff}\Sigma +O\left( c^{-3}\right) \label{eqf1} \\
&&\triangle W_{i} =-4\pi G_{eff}\Sigma ^{i}+O\left( c^{-2}\right) \label{eqf2} \\
&&\triangle W_{ij}+\partial _{i}W\partial _{j}W+2\left( 1-\beta \right) \delta
_{ij}W\triangle W =-4\pi G_{eff}\Sigma ^{ij}+O\left( c^{-1}\right) \label{eqf3}
\end{eqnarray}
while the fourth equation of (\ref{fieldeq2}) and the harmonic constraints read, using (\ref{eq:jaugeJ}) and (\ref{eq:ji})
\begin{eqnarray}
&&\triangle P+2\frac{\beta -1}{1-\gamma }W\triangle W =-4\pi G_{eff}\frac{%
\Sigma ^{kk}}{3\gamma -1}+O\left( c^{-1}\right)\\
&&\partial _{t}W+\partial _{k}W_{k}=O\left( c^{-2}\right)\\
&&\partial _{k}W_{ik}-\frac{1}{2}\partial _{i}W_{kk}+\partial _{t}W_{i}-%
\frac{1-\gamma }{2}\partial _{i}P =O\left( c^{-1}\right) \label{eq:Ji0}
\end{eqnarray}
or equivalently (after elimination of the scalar field $P$)
\begin{eqnarray}
&&\partial _{t}W+\partial _{k}W_{k}=O\left( c^{-2}\right) \label{eqf4}\\
&&\partial _{ik}W_{ik}-\frac{1}{2}\triangle W_{kk}+\partial _{ti}W_{i}+\left(
\beta -1\right) W\triangle W =-2\pi G_{eff}\frac{1-\gamma }{3\gamma -1}%
\Sigma ^{kk}+O\left( c^{-1}\right)  \label{eqf5}\\
&&\partial _{ik}W_{jk}+\partial _{ti}W_{j} =\partial _{jk}W_{ik}+\partial
_{tj}W_{i}+O\left( c^{-1}\right) . \label{eqf6}
\end{eqnarray}
The last equation refers to the fact that $\partial _{i}P$ (given by (\ref{eq:Ji0})) is
a gradient.

Note these equations are coherent with 1.5PN/BM equations assumed in \cite{KSPrD00}.

\section{Relevant field equations considering present constraints on gravity}
\label{sec:omega0}

\subsection{Without making the HG choice}

In the inner solar system, the gravitational field is such that 
\begin{equation*}
\frac{2U}{c^{2}}\sim 10^{-6}\text{ to }10^{-8}.
\end{equation*}
On the other hand, from experimental/observational constraints \cite{WLivRev06}
\begin{equation*}
\left| \gamma -1\right| \lesssim 10^{-5}\text{ ie. }\omega _{0}\gtrsim 10^{5}.
\end{equation*}
This means $ \gamma -1 $ (or $\omega _{0}^{-1}$) could be
considered numerically as a $c^{-1}$\ (at best) order quantity.\ Hence, it is convenient to present the metric under the form of a generalized
development in both powers of $c^{-1}$ and $\omega _{0}^{-1}$.\ The useful metric resulting from (\ref{jmetric}) reads (if $\omega _{0}^{\prime }$ is not
''unreasonably large'')\ 
\begin{eqnarray*}
g_{00} &=&-1+\frac{2W}{c^{2}}-\frac{2W^{2}}{c^{4}}+O\left( c^{-5},\omega
_{0}^{-1}c^{-4}\right) \\
g_{0i} &=&-\left( \gamma +1\right) \frac{2W_{i}}{c^{3}}+O\left( c^{-5}\right)
\\
g_{ij} &=&\delta _{ij}\left[ 1+\gamma \frac{2W}{c^{2}}+\frac{2W^{2}}{c^{4}}%
\right]
+\frac{4W_{ij}}{c^{4}}+O\left( c^{-5},\omega _{0}^{-1}c^{-4}\right)
\end{eqnarray*}
where $W$, $W_{i}$ and $W_{ij}$ satisfy, from (\ref{fieldeq2})
\begin{eqnarray*}
&&\square W+\frac{4}{c^{2}}\partial _{t}J =-4\pi G_{eff}\sigma +O\left(
c^{-3},\omega _{0}^{-1}c^{-2}\right) \\
&&\triangle W_{i}-\partial _{i}J =-4\pi G_{eff}\sigma ^{i}+O\left(
c^{-2}\right) \\
&&\triangle W_{ij}+\partial _{i}W\partial _{j}W-\partial _{i}J_{j}-\partial
_{j}J_{i}-2\delta _{ij}\partial _{t}J =-4\pi G_{eff}\sigma ^{ij}+O\left(
c^{-1},\omega _{0}^{-1}\right)
\end{eqnarray*}
and where $J_{i}$ reduces to 
\begin{equation}
\label{eq:NGiapprx}
J_{i}=\partial _{k}W_{ik}-\frac{1}{2}\partial _{i}W_{kk}+\partial
_{t}W_{i}+O\left( \omega _{0}^{-1}\right) .
\end{equation}
One remarks the field equations take exactly the same form as the GR case \cite{MCPRD09}
(with $G$ replaced by $G_{eff}$). The only remaining reference to
the scalar field is reduced to the presence of the $\gamma $ PN coefficient
in the metric tensor. Related to this, the field equation on $P$ is dropped out.

\subsection{Making the HG choice}

The corresponding harmonic equations to be used when considering known constraints on $\omega_0$ reads
\begin{eqnarray}
&&\square W =-4\pi G_{eff}\sigma +O\left( c^{-3},\omega
_{0}^{-1}c^{-2}\right) \label{eq:eqjH1} \notag\\
&&\triangle W_{i} =-4\pi G_{eff}\sigma ^{i}+O\left( c^{-2}\right) \label{eq:eqjH2} \\
&&\triangle W_{ij}+\partial _{i}W\partial _{j}W =-4\pi G_{eff}\sigma
^{ij}+O\left( c^{-1},\omega _{0}^{-1}\right)\label{eq:eqjH3} \notag
\end{eqnarray}

with gauge conditions 

\begin{eqnarray}
&&\partial _{t}W+\partial _{k}W_{k}=O\left( c^{-2}\right) \label{equ:h0}\\
&&\partial _{k}W_{ik}-\frac{1}{2}\partial _{i}W_{kk}+\partial \label{equ:hi}
_{t}W_{i}=O\left(c^{-1}, \omega _{0}^{-1}\right).
\end{eqnarray}

\section{Explicit harmonic metric relevant for hierarchized systems}
\label{sec:hierarchy}

\subsection{Hierarchized systems}
\label{sec:hierA}

Let us consider the case where the system is composed by bodies of masses $%
M_{A}$. Let us consider one of these bodies, named $S$, of mass $M_{S}$.\
Let us define the parameter%
\begin{equation*}
\mu =\frac{1}{M_{S}}\sum_{A\neq S}M_{A}.
\end{equation*}%
One defines a hierarchized system as a system in which the body $S$ can be chosen in such a way that%
\begin{equation*}
\mu \ll 1.
\end{equation*}
In such a system, the body S will be hereafter referred as the
"star", while the other bodies will be referred as the "planets". 

In the general relativistic $N$-body problem, multipolar moments of a body $%
A $ are defined in the coordinate system in which this body is, in some
sense, at rest. These moments are affected by coordinate transforms through a "Lorentz-like length contraction effect". These effects being of order 
$\left( u/c\right) ^{2}$, where $u$\ is the relative velocity between the
two frames, the induced effects in the metric components are of order $c^{-4}
$, since potentials are at least $c^{-2}$ terms.

In hierarchized systems, the velocity of the body $S$ is of the
order of%
\begin{equation*}
v_{S}\sim \mu v_{B}\sim \mu \sqrt{\frac{GM_{S}}{r_{S-B}}}
\end{equation*}%
where $B$ is the most
massive planet (and $r_{S-B}$\ the distance between $B$\ and the star). All the Lorentz-like contraction terms have a form like
\begin{equation*}
\frac{G M_A}{r c^2} \frac{v_A^2}{c^2}.
\end{equation*}
If $A$ is a planet ($A \neq S$), this term is at best of order $O\left( \mu c^{-4}\right)$, since $M_A \lesssim \mu M_S$. If $A$ is the star ($A = S$), this term is of order $O\left( v_S^2 c^{-4}\right)$, ie. $O\left( \mu^2 c^{-4}\right)$. Hence all these terms are, at least, of order $O\left( \mu c^{-4}\right)$.

Let us also point out that, since at this level the metric depends on time through the positions of the star and the planets only, all the terms containing the operator $\partial _{t}$ are at least of order $O(\mu)$. Hence, equations (\ref{eq:eqjH1})-(\ref{equ:hi}) lead to

\begin{eqnarray}
&&\triangle W =-4\pi G_{eff}\sigma +O\left( c^{-3},\omega
_{0}^{-1}c^{-2},\mu c^{-2}\right) \label{eq:eqjH1b} \notag\\
&&\triangle W_{i} =-4\pi G_{eff}\sigma ^{i}+O\left( c^{-2},\mu c^{-1}\right) \label{eq:eqjH2b} \\
&&\triangle W_{ij}+\partial _{i}W\partial _{j}W =-4\pi G_{eff}\sigma
^{ij}+O\left( c^{-1},\omega _{0}^{-1}, \mu\right)\label{eq:eqjH3b} \notag
\end{eqnarray}

with gauge conditions 

\begin{eqnarray}
&&\partial _{t}W+\partial _{k}W_{k}=O\left( c^{-2},\mu c^{-1}\right) \label{equ:h0b}\\
&&\partial _{k}W_{ik}-\frac{1}{2}\partial _{i}W_{kk}
=O\left(c^{-1}, \omega _{0}^{-1}, \mu\right).\label{equ:hib}
\end{eqnarray}

Related to this, $\Delta_A$ defined in \cite{SKPetAJ03} leads to numerically negligible terms (see (11.4.8) in \cite{KVPR04} for the ST version). 

\subsection{Application to the solar system}

In the Solar system, the most massive body $S$ is the Sun and one has
$$\mu \sim 10^{-3}.$$
Thus, it is legitimate to consider the Solar system as a hierarchized system. Note that, at best, only $J_2$, $J_4$ and $J_6$ planetary multipolar coefficients (for giant planets) could have a significant impact on laser ranging experiments at the required accuracy (see \cite{jacobson_jupiter,jacobson_saturn,jacobson_uranus,jacobson_neptune} for giant planets' multipole moments values). Hence, taking advantage that the Solar multipolar terms are very weak, the solution of the field equations (\ref{eq:eqjH2b}), suitable for millimetric accuracy in propagation of light, with the harmonic constraints (\ref{equ:h0b}-\ref{equ:hib}) given in barycentric coordinates turns to be

\begin{widetext}
\begin{eqnarray}
g_{00}&=&-1+\frac{2}{c^2} \left[W_0(t,\vec{x})+W_L(t,\vec{x})\right] - \frac{2 W_{S,0}^2}{c^4}+O\left( c^{-5},\omega_{0}^{-1}c^{-4},\mu c^{-4},J^S_2 c^{-4}\right) \label{eq:hugei}\\
g_{0i}&=&- 2\frac{\gamma+1}{c^3} W^i(t,\vec{x})+O\left( c^{-5},\mu c^{-4},J^S_2 c^{-4}\right) \label{eq:wifin}\\
g_{ij}&=& \left( 1 + \frac{2 \gamma}{c^2} \left[W_0(t,\vec{x})+W_L(t,\vec{x})\right]+\frac{2 W_{S,0}^2}{c^4} \right) \delta_{ij}+4 \frac{W^S_{ij}}{c^4}+O\left( c^{-5},\omega_{0}^{-1}c^{-4},\mu c^{-4},J^S_2 c^{-4}\right) \label{eq:wijfin}
\end{eqnarray}
where
\begin{eqnarray}
W_0(t,\vec{x})&=& \sum_A W_{A,0} \mbox{, and } W_{A,0} = G_{eff} \frac{M_A}{r_A(t,\vec{x})} \label{eq:Wofin}\\
W_L(t,\vec{x})&=& \sum_A W_{A,L} \mbox{, with } W_{A,L} =- G_{eff} \sum_{n=1}^{3} M_A J_{2n}^A ~~\frac{R_A^{2n}}{r_A^{2n+1}}~~ P_{2n}\left(\frac{\hat{k}_A\cdot \vec{r}_A}{r_A} \right).
\end{eqnarray}
\begin{eqnarray}
W^i(t,\vec{x})= \sum_A W^i_A(t,\vec{x})\mbox{, with }W^i_A(t,\vec{x})&=& G_{eff} \left[- \frac{\left(\vec{r}_A \times \vec{S}_A \right)^i}{2 r_A^3}+\frac{M_A v_A^i}{r_A} \left(1+\sum_{n=1}^3 J^A_{2n} \frac{R_A^{2n}}{r_A^{2n}} P_{2n} \left(\frac{\hat{k}_A \cdot \vec{r}_A}{r_A} \right) \right) \right] \label{eq:Wifin}\\
W^S_{ij}(\vec{x}) &=& \frac{1}{4} \left(G_{eff} \frac{M_S}{r_S}\right)^2 \left(\frac{(x^i-x_S^i)(x^j-x_S^j)}{r_S^2} - \delta_{ij} \right) \label{eq:hugef}
\end{eqnarray}
\end{widetext}
where one has put
$$\vec{r}_A(t,\vec{x})=\vec{x}-\vec{x}_A(t) \textrm{ and } r_A(t,\vec{x})=|\vec{r}_A(t,\vec{x})| \equiv \sqrt{(x^i-x_A^i)(x^i-x_A^i)}.$$
$M_A$, $r_a$, $v_A$ and $S_A$ are repectively the mass, the position and the velocity in barycentric coordinates, and the total angular momentum of the body $A$. $R_A$ and $J_n^A$ are the radius and the mass multipole coefficients of the body $A$. $P_n$ are the Legendre polynomials and $\hat{k}_A$ denotes the unit vector along the local $Z_A$ axis of each body $A$. 
The differences with the IAU2000 metric \cite{SKPetAJ03} lie in the presence of both the PN parameter $\gamma$ and the $c^{-4}$ space-space metric term. The multipolar term $W_L$ in $g_{ij}$ that has been neglected in the IAU2000 metric -- thanks to numerical considerations in the 1PN/BM case -- has to be considered here as well. Accordingly, multipolar terms enter also the time-space component of the metric ($g_{0i}$), and could lead to measurable effects, depending on $J_{2n}$ orders of magnitude.

While the rotational term in the time-space component of the metric is given as the usual Lense-Thirring term, slight modifications (spin multipoles) can in principle appear due to the differential rotation of the bodies. However, Solar seismology suggests \cite{corbard1} that the Sun's tachoclyne is at about $0.7$ Sun radius. Then, the mass concerned by the differential rotation is of order of a few percent of the total mass and thus, it might not lead to measurable effects. But, incidentally, time transfer and laser ranging experiments could suggest a new way (independent of Solar seismology) to test our knowledges on the solar interior dynamics. Another point coming from the solar seismology is that the solar core ($r<0.2$ Sun radius) may rotate faster than the external layers \cite{modeG1,modeG2,modeG3}. Since, the core represents a great amount of the total mass, this could affect the propagation of light at a level depending on the total angular momentum value ($\vec{S}_S$). Depending on the solar internal structure model, this may happen at the millimetric level. Using a simplified model, we show in appendix \ref{app:diffrot} how the non-rigidity of the rotation affects the metric.

Let us note that the $W_L$ terms can be neglected for inner solar system millimetric laser ranging experiments, such as Mars laser ranging for instance.

We emphasize again that neiher the $\beta$ parameter nor any $\epsilon$ parameter are required ($\epsilon$ -- corresponding sometimes to $\Lambda$ \cite{RWPRD83} or $\delta$ \cite{lator,teyssandier} -- being some PN parameter often considered in the $c^{-4}$ space-space metric term \cite{ESPRD80,KZ}). This is because both the former and the latter give too small deviations from GR to be considered in $c^{-4}$ Solar system photon's trajectories calculations. This fact is known for the former from \cite{WLivRev06} and is then obvious for the latter since it is a function of $\gamma$ and $\beta$ in the (non-massive-)ST theories considered here (as expressed in equation (\ref{jmetric})).


\section{Isotropic geodesic solutions relevant in the inner solar system}
\label{sec:prop@mmlevl}
\subsection{The metric to be used for time transfer at the millimetric level}

Space missions like LATOR \cite{lator}, TIPO \cite{tipo}, ASTROD \cite{astrod}, Phobos Laser Ranging \cite{PLR} require laser links at the millimetric level in the inner solar
system. As discussed in \cite{MCPRD09}, a full $c^{-4}$ metric like
(\ref{eq:hugei}-\ref{eq:wijfin}) is needed, and the transfer requires writting the isotropic
geodesic equations attached to this metric. However, it turns that some
terms in the list (\ref{eq:Wofin}-\ref{eq:hugef}) can be neglected. Indeed, an order of
magnitude analysis (relevant for propagation of light restricted to the inner solar
system) shows the following terms are not of numerical relevance :
\begin{itemize}
\item non-monopolar terms ($W_{L}$ and $J_{2n}^{A}$\ terms in $W^{i}$) ;
\item planetary spin terms ($\overrightarrow{S}$ terms in $W_{Planets}^{i}$).
\end{itemize}
Hence, the metric components we consider are restricted to (with $%
S_{S}^{k}=$ (constant) solar moments)%
\begin{eqnarray}
g_{00} &=&-1+\overset{(2)}{g}_{00}+\overset{(4)}{g}_{00} \notag\\
g_{0i} &=&\overset{(3)}{g}_{0i} \label{SSmetric} \\
g_{ij} &=&\delta _{ij}+\overset{(2)}{g}_{ij}+\overset{(4)}{g}_{ij} \notag
\end{eqnarray}%
with%
\begin{eqnarray*}
\overset{(2)}{g}_{00} &=&\underset{A}{\sum}\frac{2GM_{A}}{c^{2}}\frac{1}{%
r_{A}}=\underset{A}{\sum }\overset{(2)}{g}_{(A)00}  \\
\overset{(4)}{g}_{00} &=&-\frac{1}{2}\left[ \overset{(2)}{g}_{00}\right] ^{2}
\notag \\
\overset{(3)}{g}_{0i} &=&-2\left( 1+\gamma \right) \frac{G}{c^{3}}\epsilon
_{ijk}S_{S}^{j}\frac{D_{S}^{k}}{r_{S}^{3}}-\left( 1+\gamma \right) \underset{%
A}{\sum }\beta _{A}^{i}\frac{2GM_{A}}{c^{2}}\frac{1}{r_{A}}  \notag \\
\overset{(2)}{g}_{ij} &=&\gamma \delta _{ij}\overset{(2)}{g}_{00}  \notag \\
\overset{(4)}{g}_{ij} &=&-\delta _{ij}\overset{(4)}{g}_{00}+\xi _{ij}\text{
\ \ \ \ with \ \ \ \ }\xi _{ij}=\frac{G^{2}M_{S}^{2}}{c^{4}}\left( \frac{%
D_{S}^{i}D_{S}^{j}}{r_{S}^{4}}-\frac{\delta _{ij}}{r_{S}^{2}}\right)   \notag
\end{eqnarray*}%
where $D_{A}^{l}\left( t,x^{i}\right) \equiv x^{l}-x_{A}^{l}\left( t\right) $%
\ (components of $\vec{r}_A\left( t,\vec{x}\right) $%
), $r_{A}^{2}=D_{A}^{l}D_{A}^{l}$ and $\beta _{A}^{i}=v_{A}^{i}/c$.

The calculations have been made with a tracking coefficient $%
\epsilon $\ in front of $\overset{(4)}{g}_{00}$ and $\overset{(4)}{g}_{ij}$ in order to separate terms
comming from the fourth order part of the metric from those comming from
(2nd order)-(2nd order) geodesic equation coupling terms effects in the geodesic solution. However, for convenience, one has put it to unity in the following.

One gets%
\begin{eqnarray*}
\Gamma _{00}^{0} &=&\overset{(3)\text{ \ \ }}{\Gamma _{00}^{0}} \\
\Gamma _{00}^{k} &=&\overset{(2)\text{ \ \ }}{\Gamma _{00}^{k}}+\overset{(4)%
\text{ \ \ }}{\Gamma _{00}^{k}} \\
\Gamma _{0i}^{0} &=&\overset{(2)\text{ \ \ }}{\Gamma _{0i}^{0}}+\overset{(4)%
\text{ \ \ }}{\Gamma _{0i}^{0}} \\
\Gamma _{0i}^{k} &=&\overset{(3)\text{ \ \ }}{\Gamma _{0i}^{k}} \\
\Gamma _{ij}^{0} &=&\overset{(3)\text{ \ \ }}{\Gamma _{ij}^{0}} \\
\Gamma _{ij}^{k} &=&\overset{(2)\text{ \ \ }}{\Gamma _{ij}^{k}}+\overset{(4)%
\text{ \ \ }}{\Gamma _{ij}^{k}}
\end{eqnarray*}%
with, taking into account only relevant terms%
\begin{eqnarray}
\overset{(2)\text{ \ \ }}{\Gamma _{0i}^{0}} &=&\overset{(2)}{Z}_{i} \notag \\
\overset{(2)\text{ \ \ }}{\Gamma _{00}^{k}} &=&\overset{(2)}{Z}_{k} \notag \\
\overset{(2)\text{ \ \ }}{\Gamma _{ij}^{k}} &=&-\gamma \delta _{ik}\overset{%
(2)}{Z}_{j}-\gamma \delta _{jk}\overset{(2)}{Z}_{i}+\gamma \delta _{ij}%
\overset{(2)}{Z}_{k} \notag \\
\overset{(3)\text{ \ \ }}{\Gamma _{00}^{0}} &=&-\underset{(A)}{\sum }\text{ 
}\beta _{Ae}^{l}\text{ }\overset{(2)}{Z}_{(A)l} \notag \\
\overset{(3)\text{ \ \ }}{\Gamma _{ij}^{0}} &=&\overset{(3)}{Z}_{ij}+\overset%
{(3)}{Z}_{ji}+\gamma \delta _{ij}\underset{(A)}{\sum }\text{ }\beta
_{Ae}^{l}\text{ }\overset{(2)}{Z}_{(A)l} \notag \\
\overset{(3)\text{ \ \ }}{\Gamma _{0i}^{k}} &=&-\overset{(3)}{Z}_{ik}+%
\overset{(3)}{Z}_{ki}+\gamma \delta _{ik}\underset{(A)}{\sum }\text{ }\beta
_{Ae}^{l}\overset{(2)}{Z}_{(A)l} \notag \\
\overset{(4)\text{ \ \ }}{\Gamma _{0i}^{0}} &=&0 \notag \\
\overset{(4)\text{ \ \ }}{\Gamma _{00}^{k}} &=&-2\overset{(2)}{g}_{00}%
\overset{(2)}{Z}_{k} \notag \\
\overset{(4)\text{ \ \ }}{\Gamma _{ij}^{k}} &=&-\overset{(4)}{Z}_{ijk}-%
\overset{(4)}{Z}_{jik}+\overset{(4)}{Z}_{kij}. \label{eq:gammaprop}
\end{eqnarray}%
In these expressions, one has set
\begin{eqnarray*}
\overset{(2)}{Z}_{i} &\equiv &-\frac{1}{2}\partial _{i}\overset{(2)}{g}_{00}=%
\underset{(A)}{\sum }\frac{GM_{A}}{c^{2}}\frac{D_{A}^{i}}{r_{A}^{3}} \\
\overset{(3)}{Z}_{ij} &\equiv &-\frac{1}{2}\partial _{i}\overset{(3)}{g}%
_{0j}=\left( 1+\gamma \right) \frac{G}{c^{3}}\epsilon _{jmk}S_{S}^{m}\left[ 
\frac{\delta _{ik}}{r_{S}^{3}}-3\frac{D_{S}^{i}D_{S}^{k}}{r_{S}^{5}}\right]
-\left( 1+\gamma \right) \underset{(A)}{\sum }\beta _{A}^{j}\frac{GM_{A}}{%
c^{2}}\frac{D_{A}^{i}}{r_{A}^{3}} \\
\overset{(4)}{Z}_{ijk} &\equiv &-\frac{1}{2}\partial _{i}\xi _{jk}=-\frac{1}{%
2}\frac{G^{2}M_{S}^{2}}{c^{4}}\left[ \delta _{ij}\frac{D_{S}^{k}}{r_{S}^{4}}%
+\delta _{ik}\frac{D_{S}^{j}}{r_{S}^{4}}+2\delta _{jk}\frac{D_{S}^{i}}{%
r_{S}^{4}}-4\frac{D_{S}^{i}D_{S}^{j}D_{S}^{k}}{r_{S}^{6}}\right]
\end{eqnarray*}

\subsection{Isotropic geodesic solutions}
\label{sec:geodSol}

In the following, let us put%
\begin{equation*}
T\equiv x^{0}\equiv ct.
\end{equation*}%
One has now to solve the geodesic equation, written in the following (non
affine) form%
\begin{equation*}
\frac{d^{2}x^{k}}{dT^{2}}+\left( \Gamma _{\alpha \beta }^{k}-\Gamma _{\alpha
\beta }^{0}\frac{dx^{k}}{dT}\right) \frac{dx^{\alpha }}{dT}\frac{dx^{\beta }%
}{dT}=0
\end{equation*}%
with the isotropic condition%
\begin{equation*}
g_{\alpha \beta }\frac{dx^{\alpha }}{dT}\frac{dx^{\beta }}{dT}=0.
\end{equation*}%
Let us first consider the geodesic equation. It develops as%
\begin{eqnarray}
\frac{d^{2}x^{k}}{dT^{2}} &=&-\overset{(2)}{\Gamma }_{00}^{k}-\overset{(4)}{%
\Gamma }_{00}^{k}+\overset{(3)}{\Gamma }_{00}^{0}\frac{dx^{k}}{dT}-2\overset{%
(3)}{\Gamma }_{0i}^{k}\frac{dx^{i}}{dT}+2\overset{(2)}{\Gamma }_{0i}^{0}%
\frac{dx^{i}}{dT}\frac{dx^{k}}{dT} \label{eq:deveqgeod} \\
&&+2\overset{(4)}{\Gamma }_{0i}^{0}\frac{dx^{i}}{dT}\frac{dx^{k}}{dT}-%
\overset{(2)}{\Gamma }_{ij}^{k}\frac{dx^{i}}{dT}\frac{dx^{j}}{dT}-\overset{%
(4)}{\Gamma }_{ij}^{k}\frac{dx^{i}}{dT}\frac{dx^{j}}{dT}+\overset{(3)}{%
\Gamma }_{ij}^{0}\frac{dx^{i}}{dT}\frac{dx^{j}}{dT}\frac{dx^{k}}{dT} \notag
\end{eqnarray}%
The solutions can be written under the form%
\begin{equation}
x^{k}=x_{e}^{k}+N^{k}T+\overset{(2-4)}{x}^{k}  \label{soludevel}
\end{equation}%
where $\ x_{e}^{k}$\ is the
position at $t=0$ (emission). \{$N^{k}$\} is any set of three constants related by%
\begin{equation}
N^{k}N^{k}=1.  \label{norma}
\end{equation}%
The upper symbol $\left( 2-4\right) $ in $\overset{(2-4)}{x}^{k}$
means this part of the solution contains all the (numerically required) $%
c^{-2}$, $c^{-3}$ and $c^{-4}$ contributions. When useful, the notation $%
\overset{(2)}{x}^{k}$\ will represent a part of the solution that is only
required to contain all the $c^{-2}$ contributions.

Inserting the connexion (\ref{eq:gammaprop}) and (\ref{soludevel}) in the geodesic equation (\ref{eq:deveqgeod}), one gets%
\begin{eqnarray*}
\frac{d^{2}\overset{(2-4)}{x}^{k}}{dT^{2}} &=&-\left( 1+\gamma \right) 
\overset{(2)}{Z}_{k}+2\left( 1+\gamma \right) N^{k}N^{i}\overset{(2)}{Z}_{i}
\\
&&-\left( 1+\gamma \right) N^{k}\underset{(A)}{\sum }\text{ }\beta _{Ae}^{l}%
\overset{(2)}{Z}_{(A)l}+2N^{i}\left[ \overset{(3)}{Z}_{ik}-\overset{(3)}{Z}%
_{ki}+N^{j}N^{k}\overset{(3)}{Z}_{ij}\right]  \\
&&+2\overset{(2)}{g}_{00}\overset{(2)}{Z}_{k}+2N^{i}N^{j}\overset{(4)}{Z}%
_{ijk}-N^{i}N^{j}\overset{(4)}{Z}_{kij}+4N^{k}\overset{(2)}{Z}_{i}\frac{d%
\overset{(2)}{x}^{i}}{dT}+4N^{i}\overset{(2)}{Z}_{i}\frac{d\overset{(2)}{x}%
^{k}}{dT}-2\overset{(2)}{Z}_{k}N^{j}\frac{d\overset{(2)}{x}^{j}}{dT}
\end{eqnarray*}%
The expressions of $D_{A}^{l}$ to consider in this equation read, setting $D_{Ae}^{l}=x_{e}^{l}-x_{Ae}^{l}$ and\ $\beta
_{Ae}^{l}=v_{Ae}^{l}/c$ (values at $t=0$)
\begin{eqnarray*}
D_{A}^{l}\left( t,x^{i}\right)  &=&D_{Ae}^{l}+\left( N^{l}-\beta
_{Ae}^{l}\right) T+\overset{(2)}{x}^{l}\text{ \ \ \ \ in the }c^{-2}\text{
terms} \\
D_{A}^{l}\left( t,x^{i}\right)  &=&D_{Ae}^{l}+N^{l}T\text{ \ \ \ \ \ \ \ \ \
\ \ \ \ \ in the }c^{-3}\text{ and }c^{-4}\text{\ terms}
\end{eqnarray*}%
since, for a hierarchized system, all the $\beta $\ terms are at least of
order $O(\mu) $ (see section \ref{sec:hierA}). Now, let us define%
\begin{equation*}
\rho _{A}^{2}\equiv \left( T+T_{Ae}\right) ^{2}+K_{Ae}^{2}
\end{equation*}%
where (with $K_{Ae}\geq 0$)%
\begin{eqnarray}
T_{Ae} &\equiv&N^{l}D_{Ae}^{l}-\beta
_{Ae}^{l}D_{Ae}^{l}+2N^{l}D_{Ae}^{l}N^{m}\beta _{Ae}^{m}  \label{cstesdef} \\
K_{Ae}^{2} &\equiv&\left( 1+2N^{l}\beta _{Ae}^{l}\right)
D_{Ae}^{m}D_{Ae}^{m}-T_{Ae}^{2}  \notag \\
&=&\left( 1+2N^{l}\beta _{Ae}^{l}\right) D_{Ae}^{m}D_{Ae}^{m}-\left(
1+4N^{m}\beta _{Ae}^{m}\right) \left( N^{l}D_{Ae}^{l}\right) ^{2}+2\beta
_{Ae}^{m}D_{Ae}^{m}N^{l}D_{Ae}^{l}  \notag
\end{eqnarray}%
Considering only numerically relevant terms, it turns that%
\begin{equation*}
r_{A}^{2}=\left( 1-N^{l}\beta _{Ae}^{l}\right) ^{2}\left[ \rho
_{A}^{2}+2\left( D_{Ae}^{l}+N^{l}T\right) \overset{(2)}{x}^{l}\right] 
\end{equation*}%
In the $c^{-2}$ terms, one develops $D_{A}^{i}/r_{A}^{3}$ and gets%
\begin{equation}
\frac{D_{A}^{i}}{r_{A}^{3}}=\left( 1+3N^{m}\beta _{Ae}^{m}\right) \frac{%
D_{Ae}^{i}+\left( N^{i}-\beta _{Ae}^{i}\right) T}{\rho _{A}^{3}}+\frac{%
\overset{(2)}{x}^{i}}{\rho _{A}^{3}}-3\left( D_{Ae}^{l}+N^{l}T\right) 
\overset{(2)}{x}^{l}\frac{D_{Ae}^{i}+N^{i}T}{\rho _{A}^{5}}.
\label{forcedevel}
\end{equation}%
This way, the geodesic's equation now reads, discarding non-relevant terms%
\begin{eqnarray}
\frac{d^{2}\overset{(2-4)}{x}^{k}}{dT^{2}} &=&-\left( 1+\gamma \right)
\left( \delta _{ik}-2N^{i}N^{k}\right) \underset{(A)}{\sum }\frac{GM_{A}}{%
c^{2}}\left( 1+3N^{m}\beta _{Ae}^{m}\right) \frac{D_{Ae}^{i}+\left(
N^{i}-\beta _{Ae}^{i}\right) T}{\rho _{A}^{3}}  \label{geodezeq} \\
&&-\left( 1+\gamma \right) \underset{(A)}{\sum }\frac{GM_{A}}{c^{2}}\left(
N^{k}\frac{\beta _{Ae}^{l}D_{A}^{l}}{\rho _{A}^{3}}+2\beta _{A}^{k}\frac{%
N^{i}D_{A}^{i}}{\rho _{A}^{3}}-2N^{i}\beta _{A}^{i}\frac{D_{A}^{k}}{\rho
_{A}^{3}}+2N^{k}N^{j}\beta _{A}^{j}\frac{N^{i}D_{A}^{i}}{\rho _{A}^{3}}%
\right)   \notag \\
&&-\left( 1+\gamma \right) \frac{G}{c^{3}}S_{S}^{m}\left[ 4N^{i}\epsilon
_{imk}\frac{1}{\rho _{S}^{3}}+6\epsilon _{kml}\frac{N^{i}D_{S}^{i}D_{S}^{l}}{%
\rho _{S}^{5}}-6N^{i}\epsilon _{iml}\frac{D_{S}^{k}D_{S}^{l}}{\rho _{S}^{5}}%
+6N^{j}N^{k}\epsilon _{jml}\frac{N^{i}D_{S}^{i}D_{S}^{l}}{\rho _{S}^{5}}%
\right]   \notag \\
&&-\frac{G^{2}M_{S}^{2}}{c^{4}}\left[ -4\frac{D_{S}^{k}}{\rho _{S}^{4}}%
+2N^{k}\frac{N^{i}D_{S}^{i}}{\rho _{S}^{4}}-2\frac{\left(
N^{i}D_{S}^{i}\right) ^{2}D_{S}^{k}}{\rho _{S}^{6}}\right]   \notag \\
&&-\frac{GM_{S}}{c^{2}}\left[ -4N^{k}\frac{D_{S}^{i}}{\rho _{S}^{3}}\frac{d%
\overset{(2)}{x}^{i}}{dT}-4\frac{N^{i}D_{S}^{i}}{\rho _{S}^{3}}\frac{d%
\overset{(2)}{x}^{k}}{dT}+2\frac{D_{S}^{k}}{\rho _{S}^{3}}N^{i}\frac{d%
\overset{(2)}{x}^{i}}{dT}\right]   \notag \\
&&-2\left( \delta _{ik}-2N^{i}N^{k}\right) \frac{GM_{S}}{c^{2}}\left[ \frac{%
\overset{(2)}{x}^{i}}{\rho _{S}^{3}}-3\left( D_{Se}^{m}+N^{m}T\right) 
\overset{(2)}{x}^{m}\frac{D_{Se}^{i}+N^{i}T}{\rho _{S}^{5}}\right] .  \notag
\end{eqnarray}%
Let us point out the development (\ref{forcedevel}) is justified only if one has%
\begin{equation*}
\frac{15}{2}\left[ \frac{\left( D_{Ae}^{l}+N^{l}T\right) \overset{(2)}{x}^{l}%
}{\rho _{A}^{2}}\right] ^{2} \ll 3\frac{\left( D_{Ae}^{l}+N^{l}T\right) \overset%
{(2)}{x}^{l}}{\rho _{A}^{2}}
\end{equation*}%
on the whole photon's orbit. In fact, since only the solar terms are
required for the higest order contributing terms, we consider this condition with respect to the Sun only. Besides, considering applications aimed in this paper, for which the contribution of
the $c^{-4}$ terms is expected to be close to the limit of detectability, it
is sufficient to require%
\begin{equation}
\left\vert \left( D_{Se}^{l}+N^{l}T\right) \overset{(2)}{x}^{l}\right\vert
\lesssim \frac{2}{5}\rho _{S}^{2}\times f  \label{develcond}
\end{equation}%
where $f\sim 1/10$, or even $1/100$, to keep a safety margin.

One first obtains the solution for the $c^{-2}$ and $c^{-3}$ contributions.\
Inserting the $c^{-2}$ solution in the rhs of (\ref{geodezeq}), one
gets the $c^{-4}$ contribution.\ Since one considers isotropic
geodesics only, the solution has also to satisfy the isotropic condition, that
writes, using (\ref{SSmetric}) and (\ref{soludevel}) (and writting $\gamma =1
$\ in the $c^{-4}$\ terms)%

\begin{equation*}
\left( 1+\gamma \right) \overset{(2)}{g}_{00}+2N^{i}\overset{(3)}{g}%
_{0i}+2N^{i}\frac{d\overset{(2-4)}{x}^{i}}{dT}+2\overset{(2)}{g}_{00}N^{i}%
\frac{d\overset{(2)}{x}^{i}}{dT}+\frac{d\overset{(2)}{x}^{i}}{dT}\frac{d%
\overset{(2)}{x}^{i}}{dT}+N^{i}N^{j}\xi _{ij}=0
\end{equation*}%
Finally, after some tedious calculations, one gets the solution%
\begin{eqnarray}
x^{k}\left( T\right)  &=&x_{e}^{k}+\left( 1-2\frac{G^{2}M_{S}^{2}}{c^{4}}%
\frac{1}{K_{Se}^{2}}\right) N^{k}T+\left( 1+\gamma \right) \underset{(A)}{%
\sum }\frac{GM_{A}}{c^{2}}\left[ \overset{(2)}{X}_{A}^{k}\left( T\right) -%
\overset{(2)}{X}_{A}^{k}\left( T=0\right) \right]   \label{geodesol} \\
&&+\left( 1+\gamma \right) \underset{(A)}{\sum }\frac{GM_{A}}{c^{2}}\left[ 
\overset{(3)}{Y}_{A}^{k}\left( T\right) -\overset{(3)}{Y}_{A}^{k}\left(
T=0\right) \right] +\left( 1+\gamma \right) \frac{G}{c^{3}}S_{S}^{m}\left[ 
\overset{(3)}{Z}^{km}\left( T\right) -\overset{(3)}{Z}^{km}\left( T=0\right) %
\right]   \notag \\
&&+\frac{G^{2}M_{S}^{2}}{c^{4}}\left[ \overset{(4)}{X}^{k}\left( T\right) -%
\overset{(4)}{X}^{k}\left( T=0\right) \right]   \notag
\end{eqnarray}%
where one has set%
\begin{eqnarray*}
\overset{(2)}{X}_{A}^{k} &=&-\frac{D_{Ae}^{k}-N^{k}N^{l}D_{Ae}^{l}}{%
K_{Ae}^{2}}\rho _{A}-N^{k}\ln \left( \rho _{A}+T+T_{Ae}\right)  \\
\overset{(3)}{Y}_{A}^{k} &=&-\frac{\left( \delta ^{km}-N^{k}N^{m}\right)
\left( \beta _{Ae}^{l}D_{Ae}^{m}+D_{Ae}^{l}\beta _{Ae}^{m}\right) N^{l}}{%
K_{Ae}^{2}}\rho _{A}+\beta _{Ae}^{k}\ln \left( \rho _{A}+T+T_{Ae}\right)  \\
\overset{(3)}{Z}^{km} &=&2\frac{D_{Se}^{k}-N^{k}N^{l}D_{Ae}^{l}}{K_{Se}^{2}}%
N^{i}\epsilon _{imn}\left( 2\frac{D_{Se}^{n}}{K_{Se}^{2}}\rho _{S}-D_{Se}^{n}%
\frac{1}{\rho _{S}}\right)  \\
&&+2\epsilon _{kmi}\left[ N^{i}\left( \frac{1}{K_{Se}^{2}}\rho _{S}-\frac{1}{%
\rho _{S}}\right) +\frac{D_{Se}^{i}-N^{i}N^{l}D_{Se}^{l}}{K_{Se}^{2}}\frac{%
T+T_{Se}}{\rho _{S}}\right]  \\
\overset{(4)}{X}^{k} &=&\frac{D_{Se}^{k}-N^{k}N^{l}D_{Se}^{l}}{K_{Se}^{2}}%
\left[ 4\frac{T+T_{Se}}{\rho _{S}}\ln \frac{\rho _{S}+T+T_{Se}}{\rho
_{Se}+T_{Se}}-\left(4-\frac{\epsilon}{4} \right)\frac{T+T_{Se}}{K_{Se}}\arctan \frac{T+T_{Se}}{%
K_{Se}}+\frac{\epsilon}{4}\frac{K_{Se}^{2}}{\rho _{S}^{2}}-\frac{4\rho _{Se}}{\rho _{S}}+%
\frac{4\rho _{Se}\rho _{S}}{K_{Se}^{2}}\right]  \\
&&+\frac{N^{k}}{K_{Se}}\left[ \frac{4K_{Se}}{\rho _{S}}\ln \frac{\rho
_{S}+T+T_{Se}}{\rho _{Se}+T_{Se}}-\left(4-\frac{\epsilon}{4} \right)\arctan \frac{T+T_{Se}}{K_{Se}}%
+\frac{\epsilon}{4}K_{Se}\frac{T+T_{Se}}{\rho _{S}^{2}}+4\frac{\rho _{Se}}{K_{Se}}%
\frac{T+T_{Se}}{\rho _{S}}\right] 
\end{eqnarray*}%
and with $\rho _{Ae}^{2}\equiv \rho _{A}^{2}\left( T=0\right)
=T_{Ae}^{2}+K_{Ae}^{2}=\left( 1+2N^{l}\beta _{Ae}^{l}\right)
D_{Ae}^{m}D_{Ae}^{m}$. The last expression is given with the tracking coefficient $\epsilon$ (=1 in applications) in order to isolate terms coming from the $c^{-4}$ part of the metric in the $c^{-4}$ part of the geodesic solution.

The solution (\ref{geodesol}) has been written in such a way that :
\begin{itemize}
\item all the velocity and spin source terms have been brought together in the $%
c^{-3}$ contribution (apart from the $\beta $\ terms entering the
definitions of $T_{Ae}$\ and $K_{Ae}$ in (\ref{cstesdef}));
\item terms parallel to $N$ and orthogonal to $N$ (i.e. $\delta ^{km}-N^{k}N^{m}$%
, $D_{Ae}^{k}-N^{k}N^{l}D_{Ae}^{l}$ and $\epsilon _{kmi}N^{i}$) appear
explicitely.
\end{itemize}
Let us also point out that :
\begin{itemize}
\item thanks to the hierarchy, it is sufficient to put%
\begin{eqnarray}
T_{Ae} &=&N^{l}D_{Ae}^{l}  \label{cstesapprox} \\
K_{Ae}^{2} &=&D_{Ae}^{m}D_{Ae}^{m}-T_{Ae}^{2}=D_{Ae}^{m}D_{Ae}^{m}-\left(
N^{l}D_{Ae}^{l}\right) ^{2}  \notag
\end{eqnarray}%
in $\overset{(3)}{Y}_{A}^{k}$, $\overset{(3)}{Z}^{km}$ and $\overset{(4)}{X}%
^{k}$, instead of the full definitions (\ref{cstesdef});
\item the $\beta \ln \left( ...\right) $ term in $\overset{(3)}{Y}_{A}^{k}$ and $%
\epsilon _{kmi}S_{S}^{m}D_{Se}^{i}$ term entering $\overset{(3)}{Z}%
^{km}S_{S}^{m}$\ are the only terms that are neither parallel nor orthogonal
to $N$.
\end{itemize}
For geometrical interpretations, let us remark that :
\begin{itemize}
\item in some sense (among others, since $K_{Ae}$ is defined from the initial
position, and not from the (virtual) position at $t=-\infty $, see the
discussion in \cite{KZ}), $K_{Ae}$ can be interpreted (from (\ref{cstesapprox})) as some
impact parameter, or minimal approach
distance, with respect to the body $A$;
\item to the same extent, the quantity $\left(
D_{Ae}^{k}-N^{k}N^{l}D_{Ae}^{l}\right) /K_{Ae}^{2}$ can be interpreted as a
vector of modulous $1/K_{Ae}$\ that is pointing toward the closest approach
point (with respect to the body $A$) ;
\item to the same extent, the quantity $T_{Ae}$ can be interpreted as the
opposite of time needed to reach the point of closest approach to the body $A
$\ from the initial position $x_{e}^{k}=x^{k}\left( T=0\right) .$
\end{itemize}
\subsection{Discussion}
\label{sec:discuss}

In the inner Solar system, the equation (\ref{geodesol}) gives the
time-dependent position for a photon starting from the position $x_{e}^{k}$\
at $t=0$. The millimetric level expected in forthcoming experiments has
required the post-post-newtonian level of approximation.\ The trajectory is
characterized by two arbitrary parameters, giving the initial "direction" in
some sense.\ These parameters are, for instance, two any components of the
"vector" $N^{k}$, the third being given by the normalization condition
(\ref{norma}).\ (In relevant applications, the requirement the photon
has to reach a given target having a known orbit, for instance, fixes these
two remaining degrees of freedom and the coordinate time of reception.)

So far, such post-post-newtonian works have been completed in some one gravitationnal source cases :
\begin{itemize}
\item spherical source in GR \cite{livrebrumberg};
\item non-spherical source in GR \cite{RWPRD83};
\item spherical source in a class of parametrized metric, including GR and, to some extent, ST theories \cite{KZ}.
\end{itemize}
The current work includes in a coherent way the different gravity sources of the solar system
at the required numerical level.\ This includes kinetic source terms effects
(velocities of the sources, rotation of the Sun), leading to non-zero
time-space components of the metric ($g_{0i}$), and to the related $c^{-3}$
contributions in the solution (\ref{geodesol}).

However, it has been stressed that (\ref{develcond}) has to be
satisfied on the whole trajectory to ensure the validity of the proposed
geodesic solution. One has pointed $K_{Se}$ can be crudely interpreted as
the closest approach distance to the Sun $\rho _{S,min}$ on the photon's orbit,
and $T_{Se}$ as the opposite of time needed to reach the point of closest
approach to the Sun from the initial position. The condition
(\ref{develcond}) leads to%
\begin{equation*}
\frac{2GM_{S}}{c^{2}}\left\vert \rho _{S}-\rho _{Se}+\left( T+T_{Se}\right)
\ln \frac{T+T_{Se}+\rho _{S}}{T_{Se}+\rho _{Se}}\right\vert \lesssim \frac{2%
}{5}\rho _{S}^{2}\times f.
\end{equation*}%
For a given orbit (thence a given $\rho _{S,min}$), and if $\rho _{Se} \gg \rho
_{Sm}$, the most severe condition is reached at the closest approach,
leading to%
\begin{equation}
\rho _{Se}\lesssim \rho _{S,min}\times \left( \frac{2GM_{S}}{\rho _{S,min}c^{2}}%
\right) ^{-1}\times \frac{2f}{5}.  \label{develcond2}
\end{equation}%
Since $\rho _{S,min}$ cannot be smaller than the solar radius, it turns the
validity of the proposed geodesic solution is ensured as soon as%
\begin{equation*}
\rho _{Se}\lesssim \left( 7.10^{5}\ \ km\right) \times \left( \frac{3\ km}{%
7.10^{5}\ km}\right) ^{-1}\times \frac{2f}{5}~~\sim~~ 10^{9}\ \ km
\end{equation*}%
the last estimation being obtained taking $2f/3\sim 1/100$. This ensures the
validity of the proposed solution for inner solar system links.

While it could seem natural to claim that any position on the photon's orbit can
be regarded as the initial one, a glance at the formula
(\ref{geodesol}) makes obvious this cannot be the case.\ Indeed, at
great "distance" (i.e. $T\longrightarrow \pm \infty $), one has%
\begin{equation}
\frac{d\overset{(4)}{X}^{k}}{dT}=s_{T}\frac{D_{Se}^{k}-N^{k}N^{l}D_{Se}^{l}}{%
K_{Se}^{2}}\left[ -\left(4-\frac{\epsilon}{4}\right)\frac{1}{K_{Se}}\frac{%
\pi }{2}+\frac{4\rho _{Se}}{K_{Se}^{2}}\right] +\text{\ (terms that }%
\longrightarrow 0\text{ when }\left\vert T\right\vert \longrightarrow \infty 
\text{)}  \label{enhancedterm}
\end{equation}%
where $s_{T}$ is the sign of $T$. On the other hand, it is clear on physical grounds that $dx^{k}/dT$ should
go to values that don't depend on the chosen initial point on the orbit when $T\longrightarrow \pm \infty $, these limits
corresponding to "initial" and "final" directions of the photon (the
difference of these limits giving the so-called "deviation" of the photon induced
by the gravitational field). It is clear from (\ref{enhancedterm}) that 
$dx^{k}/dT$ computed from (\ref{geodesol}) includes a $c^{-4}$ term
that goes to infinity for an initial position chosen (on any given orbit)
arbitrarily far from the Sun.\ This is obviously physically unacceptable
(and incompatible with the fact that $c^{-4}$ terms should be corrective
terms). This term corresponds to one of the so-called "enhanced terms" in \cite{
KZ} (also present in \cite{livrebrumberg}). The validity condition (\ref{develcond}), rewritten
(\ref{develcond2}), prevents to fall in this case.\ The tracking coefficient $\epsilon $\ shows the involved term
comes from coupling contributions in the geodesic equation
(\ref{geodezeq}) (i.e from (\ref{geodezeq})-r.h.s. terms
depending explicitely on $\overset{(2)}{x}^{i}$), and not from the fourth
order terms $\overset{(4)}{g}_{\alpha \beta}$\ in the metric, in accordance with \cite{KZ}.

Since the validity condition is ensured only for initial conditions at
distances not exceding some astronomical units, this shows one should be
careful in any extention of such analytical post-post-newtonian approaches
to both the deviation of light and light tranfer problems at the external
solar system scale, at least if photons grazing the Sun are considered
(remark it is not the case for the GAIA\ mission). 

\section{Conclusion}

In this paper, one has get the explicit geodesic equation (\ref{geodesol}) relevant
for propagation of light at the millimetric level in the inner Solar system,
and for related problems. These expressions should be useful for experiments like \cite{lator,tipo,astrod,PLR}. One has also explicitly given the condition of validity
of this expression, and pointed out how this validity condition is related
to the method used to solve the geodesic equation at the post-post-newtonian
level.

This work can be extended to missions involving links at the whole Solar
system scale.\ The planetary $c^{-4}$ metric contributing terms are expected
to be negligible in this case too, since these terms generate effects of order of $%
10^{-16}\ s$ at best, i.e. well under the mm level, even with a large safety margin.
Hence, the metric (\ref{eq:hugei}-\ref{eq:wijfin}) is relevant in this case too.\ However,
giant planets multipolar terms have to be taken into account in (\ref{eq:Wofin}-\ref{eq:hugef}),
at least at the $c^{-2}$\ level, as shown by previous studies (see \cite{CM-CQG} for instance).
Besides, the validity condition of the method has to be carefully checked in
this case, especially when one deals with photons grazing the Sun
before/after travelling the external Solar system as argued in section \ref{sec:discuss}. 

\begin{acknowledgments}
Olivier Minazzoli wants to thank the Government of the Principality of Monaco for their financial support.
\end{acknowledgments}

\appendix
\section{Effects on the metric of a non-rigid rotation}
\label{app:diffrot}

Recent results coming from Solar seismology suggest that the Solar core rotates faster than the external layers \cite{modeG1,modeG2,modeG3}. In the following toy model we consider the mass density as spherical ($\rho(\vec{x})=\rho(r)$). As usual in Solar models \cite{corbard1}, we make the distinction between three main regions : the core region ($r \in [0,R_C \approx 0,2 R_S]$, where $R_C$ is the core radius), the radiative region ($r \in [R_C,R_R\approx 0,7 R_S]$, where $R_R$ is the tachoclyne radius) and the convection region ($r \in [R_R,R_S]$, where $R_S$ is the Sun radius). $\vec{\Omega}$ being the angular velocity, we modelize the differential rotation as follows
\begin{equation}
\vec{\Omega}(\vec{x})=\Omega(r) \hat{k}_z,
\end{equation}
with
\begin{eqnarray}
\Omega(r)= \Omega_C(\theta)~ \Pi \left(\frac{r}{R_c}\right) 
+\Omega_R(\theta)~ \Pi \left(\frac{r-R_C}{R_R-R_C}\right) 
+\Omega_D(\theta)~ \Pi \left(\frac{r-R_R}{R_D-R_R}\right) ,
\end{eqnarray}
where 
\begin{eqnarray*}
\Pi(x) &=& 1 ~\forall~ x \in [0;1]\\
&=&0~ \mbox{ everywhere else},
\end{eqnarray*}
and where we modelize the differential rotation by a simple model in accordance with the usual model \cite{corbard2}
\begin{equation}
\Omega_B(\theta) = \Omega_B \left(1+\epsilon_B~ cos^2 \theta \right),
\end{equation}
where $\Omega_B$ and $\epsilon_B$ are constants, $B$ being any of the three previous regions ($C$, $R$ or $D$).

Solar seismology suggests that the radiative region rotates as a solid -- meaning $\Omega_R(\theta)=\Omega_R$ (ie. $\epsilon_R=0$). However, since we are interested in testing solar seismology results, let us relax this assumption. Let us write
\begin{equation}
\bar{W}^i_S(t,\vec{x})=\bar{W}^i_C(t,\vec{x})+\bar{W}^i_R(t,\vec{x})+\bar{W}^i_D(t,\vec{x}),
\end{equation}
with
\begin{equation}
\bar{W}^i_B(t,\vec{x})=G \int_B d^3X \left(\Omega_B(R) \times \vec{X} \right)^i \frac{\rho(R)}{|\vec{x}-\vec{X}|}.
\end{equation}
$\bar{W}^i_S$ being the Sun spin part of $W_i$ in (\ref{eq:wifin}). Then the solution writes
\begin{eqnarray}
&& \bar{W}^i_S(t,\vec{x}) =4 \pi G \sum_B \label{appeq:lens++} \left(\vec{\Omega}_B \times \vec{x} \right)^i \left[\left(\frac{1}{3}+\frac{\epsilon_B}{15} \right) \frac{M_2^B}{r^3} + \frac{\epsilon_B}{35} \frac{M_4^B}{r^5} (5~cos^2 \theta-1) \right],
\end{eqnarray}
where
\begin{eqnarray*}
M_N^C = M_N(0,R_C),~~ M_N^R = M_N(R_C,R_R),~~
M_N^D = M_N(R_R,R_S),
\end{eqnarray*}
with
\begin{equation}
M_N(X,Y)=\int_X^Y R^{N+2} \rho(R) dR.
\end{equation}

First, note that a faster rotation of a rigid core will modify the value of the total angular momentum only. However, this value could be affected by a differential rotation as well.

But, one also may have to consider a term like the last term of the r.h.s. of (\ref{appeq:lens++}) in (\ref{eq:Wifin}), in order to measure possible weak effects due to differential rotations of the different stages of the Sun -- then, giving a characteristic way to put constraints on such differential rotations, which will be independent of Solar seismology and neutrino detection results. In what follows, we will refer to this term as the differential-Lense-Thirring term.

As an illustration, let us consider the following toy model. Assume (1) the density decreases linearly with the distance to the center of the Sun (2) the differential rotation is independent of the distance from the center. Now consider that the photons -- used for the time transfer -- graze the Sun (ie. $b=\alpha R_S$, where $b$ is the impact parameter and $\alpha(>1)$ a parameter ideally close to 1). Then the effect of the differential-Lense-Thirring term is about $\alpha^{-2}\epsilon_B/11$ times the usual Lense-Thirring effect.

However, realistic models of the Sun and its rotation are expected to substantially decrease this effect. A specific study, that considers different realistic models of the Sun, should be done in order to clarify this issue.

\end{document}